\documentclass[%
reprint,
superscriptaddress,
 amsmath,amssymb,
 aps,
 prl,
 lengthcheck,%
]{revtex4}
\pdfoutput=1

\usepackage{mathptmx}

\usepackage[squaren]{SIunits}  
\usepackage{graphicx}    
\usepackage{color}

\usepackage{hyperref}

\begin{document}


\title{{\color{black} Fully fault tolerant quantum computation with non-deterministic gates}}




\author{Ying Li}
\affiliation{Centre for Quantum Technologies, National University of Singapore, 3 Science Drive 2, Singapore 117543}
\author{Sean D. Barrett}
\affiliation{Blackett Laboratory and Institute for Mathematical Sciences, Imperial College London, London SW7 2PG, United Kingdom}
\author{Thomas M. Stace}
\affiliation{School of Mathematics and Physics, University of
Queensland, Brisbane, QLD 4072, Australia}
\affiliation{Centre for Quantum Technologies, National University of Singapore, 3 Science Drive 2, Singapore 117543}
\author{Simon C. Benjamin}
\affiliation{Department of Materials, University of Oxford, Parks Road, Oxford OX1 3PH, UK}
\affiliation{Centre for Quantum Technologies, National University of Singapore, 3 Science Drive 2, Singapore 117543}

\date{\today}

\begin{abstract} In certain approaches to quantum computing the operations between qubits are non-deterministic and likely to fail. For example, a distributed quantum processor would achieve scalability by networking together many small components; operations between components should assumed to be failure prone. In the logical limit of this architecture each component contains only one qubit. Here we derive thresholds for fault tolerant quantum computation under such extreme paradigms. We find that computation is supported for remarkably high failure rates (exceeding $90\%$) providing that failures are heralded, meanwhile the rate of unknown errors should not exceed $2$ in $10^{4}$ operations. 
\end{abstract}


\maketitle

The field of quantum information processing (QIP) has seen many experimental successes, but the challenge of scaling from a few qubits to large scale devices remains unsolved. One can argue that the issue is so crucial that it should dictate the choice of fundamental architecture for the machine. For example, in the concept of {\em distributed} QIP a plurality of small components, each similar in complexity to systems already realised experimentally, are networked together to constitute a full scale machine. The components may be trapped atoms or ions, or solid state nanostructures such as quantum dots or NV centres~\cite{Benjamin:2009p374}. Each component can be presumed to be under good control, and it is understood that the key task is then to entangle the physically remote components.
 {\color{black} An attractive method of achieving this entangling operation (EO) is to arrange for each component to emit a photon that is correlated with the internal state of the component, before performing a joint measurement (with the aid of simple linear optical elements) of the photons.} A considerable number of such entanglement schemes have been advanced since the first ideas in 1999~\cite{Cabrillo:1999p339,Bose:1999p326}. An important step was the realisation that photon loss can be detected, or heralded, within such a protocol ~\cite{Barrett:2005p363,Lim:2005p364}. Generally in these remote entanglement protocols, one is supposed to employ optical measurements that simultaneously observe two, or even four~\cite{Benjamin:2005p362}, components simultaneously. This principle for generating entanglement has in fact been demonstrated experimentally: first with ensemble systems~\cite{Chou:2005p333} and subsequently with individual atoms~\cite{Moehring:2007p337}.

It is understood that the remote EOs may be failure prone. However, these failures are assumed to be {\em heralded}: the experimentalist is aware when a failure occurs. The appropriate strategy for dealing with such failures depends
on the level of complexity within each component. In the case that each component incorporates
multiple qubits then we can nominate one `logical qubit' and use the other(s) to make repeated attempts at remote entanglement; when we are eventually successful then we can transfer the entanglement to the logical qubits~\cite{BriegelDur03,Benjamin:2006p358}. However, many
physical systems may have only very limited complexity, and moreover it is always desirable to minimise the required complexity. Therefore it is interesting to consider the case of just one qubit in each component. This may be thought of as the extreme limit of the distributed paradigm. If we suppose that the probability $p_h$ of a heralded error is high, perhaps well above $50\%$, then it is clear that we cannot perform quantum computation by directly implementing a standard circuit model approach.  {\color{black} However, it has been shown that even in spite of such heralded failures, arbitrary quantum algorithms can be implemented ~\cite{Lim:2005p364,Barrett:2005p363,BK_comment,Nielsen:2004p371,Duan:2005p369,Rohde:2007p370, UTprl}.} These insights are related to earlier ideas on photonic QIP~\cite{YoranAndReznik, BrowneAndRudolph}. {\color{black} While such schemes demonstrated that large heralded failure rates can be tolerated, this was not shown in a fully fault-tolerant manner. In particular, it was not known if large heralded failure rates can be tolerated in the presence of realistic error rates for all other elementary operations.}

Fortunately, other studies have developed an approach which can be adapted to present purposes. Recently a series of beautiful results by Raussendorf, Harrington and others described a method for QIP which involves creating a large scale cluster state with a regular three dimensional lattice structure  \cite{PhysRevLett.98.190504,Raussendorf20062242,1367-2630-9-6-199}.  Defect regions  within the 3D lattice are braided together, yielding topologically protected clifford gates. QIP implemented using this topologically protected cluster (TPC) state has a remarkably large tolerance against elementary errors (at rates $\lesssim1\%$) during preparation, entangling operations and single qubit measurement. Subsequently, two of us have extended this idea to incorporate the possibility that the lattice contains a significant proportion of missing qubits at  known locations (nearly $25\%$ can be missing) \cite{PhysRevLett.102.200501,barrett2010fault,PhysRevA.81.022317}. 

Here we consider the generation of  a TPC state when the entangling operations are themselves subject to heralded failures during the cluster state growth process. The result is a lattice with a certain proportion of known failed entanglement relations (missing `edges' in the graph state). The task of determining a threshold for universal QIP depends on proper choice of growth strategy together with a careful audit of the accumulation of unknown errors in that process. We show how to map this cluster state with missing `edges' to one with missing qubits, thereby making contact with the loss-tolerant thresholds quoted in the prior literature \cite{PhysRevLett.102.200501,barrett2010fault,PhysRevA.81.022317}. 

\begin{figure}
  \centering
  \includegraphics[width=8cm]{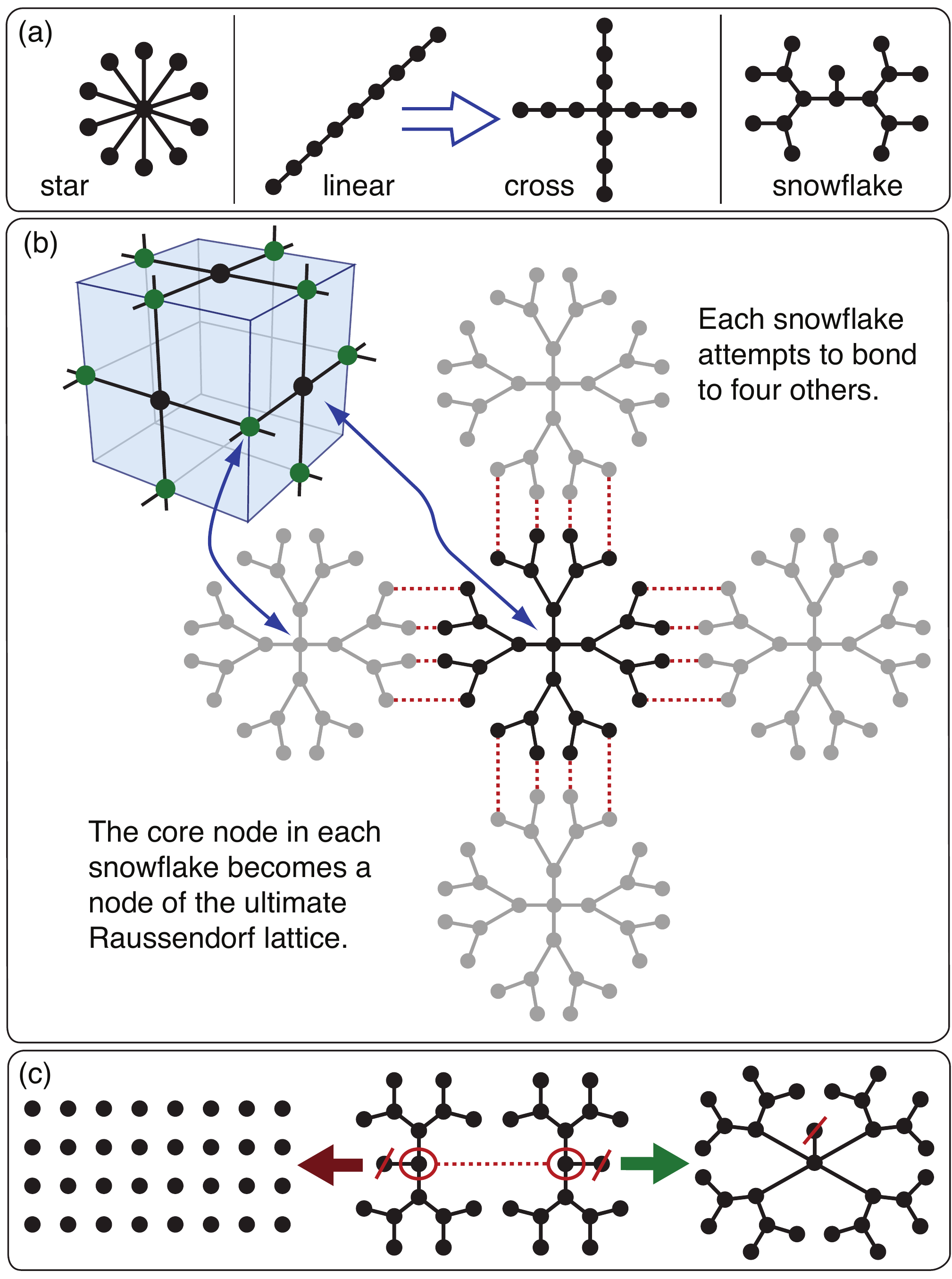}
 \caption{The figure shows {\em graph states}: nodes correspond to qubits, and connections (`edges') correspond to phase entanglement. (a) The `building block' resources that have been considered by previous authors. (b) Illustration of how one would synthesise the TPC state (depicted in upper left) by fusing together snowflake resource objects. One would use a quarter of the structure to bond with each of four adjacent objects; basically the same approach applies for the star or the cross geometries. After the bonding stage, there must be a `pruning' phase where we remove all but the core nodes and thus simplify down to the target lattice. (c) One minor revision in the case of the snowflake is that the final round involves fusing qubits that are one step from the `core' in order to generate a 4-node.  }
  \label{fig:topologies}
\end{figure}

Several previous papers have considered the task of creating large entangled states when the elementary EO is failure prone (see Fig.~\ref{fig:topologies} and caption). In principle 
a `divide and conquer' approach can permit the entangled state to have positive growth
{\em on average} for any \textcolor{black}{nonzero success probability $p_s = 1-p_h$}~\cite{Lim:2005p364,Barrett:2005p363,BK_comment,Nielsen:2004p371,Duan:2005p369,Rohde:2007p370, UTprl}. Generally the solution involves generating relatively small resource states and subsequently connecting them. As shown in Fig.~\ref{fig:topologies}(a) the possible `building block' resources include stars~\cite{Nielsen:2004p371}, linear clusters~\cite{Barrett:2005p363,BK_comment} which in turn give rise to cross structures~\cite{Duan:2005p369}, and tree topologies~\cite{BodiyaPRL06}. The last of these, also called the `snowflake', has been proposed as a optimal choice for minimising errors~\cite{UTprl}.

In the present paper our aim is to synthesise the TPC state (Fig.~\ref{fig:topologies}b, inset top left). This structure has the property that each node has four neighbours. Therefore we attempt to entangle together each resource with four others, as depicted in Fig.~\ref{fig:topologies}b). In the particular example illustrated we see that there will be $N=4$ attempts to connect to each of the surrounding snowflakes. If one or more of these attempts succeeds, then have successfully connected the snowflakes, while with probability $p_h^N$ all attempts will fail and the resulting TPC state will have a missing `edge' at that point. These missing edges are known, and therefore are not errors but rather defects which we must allow for in the subsequent computation. Obviously, it will be necessary to create resource objects which are sufficiently large so that this net failure probability is below the threshold for fault tolerant QIP, which we presently discuss. For high values of $p_h$ we will see that the resource states must be considerably larger than those illustrated in Fig.~\ref{fig:topologies}.

{\color{black} In order to evaluate this scheme, }it is essential to {\color{black} determine} the accumulation of {\em unknown} errors when we perform the star, cross, and snowflake strategies. These errors will occur during growth of the resource, during the fusion of resources (as in Fig.~\ref{fig:topologies}b), and also during the process of removing redundant qubits to simplify down to the TPC state. In order to minimise error accumulation during growth we make the aggressive choice that {\em whenever} there is a known failure during the growth of the resource object, that entire resource object is abandoned. Fortunately all three of the resources we consider -- star, cross, snowflake -- can be grown through a series of steps each of which (on success) doubles the entity's size. Thus the process is quick in the sense that it requires a number of successful steps that is merely a logarithmic function of the target resource size.  Here we assume each resource is grown using one of two forms of EO: we use either parity projections, i.e. projecting a pair of qubits into the odd or even parity subspaces, or a canonical control-phase gate between the qubit pair, depending on which is more efficient~\cite{supMat}. Both operations are known to be possible through suitable measurements on emitted photons~\cite{Lim:2005p364,Barrett:2005p363,BK_comment}.

We consider various forms of error. Single-qubit errors may occur during preparation, or while performing a single-qubit
rotation, or during measurement. Moreover these errors may also occur passively in memory, i.e. there is a rate at which qubits decohere even when not part of any active operation. Meanwhile two-qubit errors may occur when we perform entanglement operations. We account for imperfections both in the emission of photons (e.g. from an imperfect selection rule in an atomic system, say) and errors arising from imperfect measurement of emitted photons~\cite{supMat}. Different probabilities are assigned to the various errors, however, for simplicity in generating the diagrams here we set the rates for all forms of the active `gate' errors to be equal and we denote their probability $p_G$. Memory errors are considered separately later in the paper.

A principle conclusion from our analysis of error propagation is that two-qubit errors occurring during the growth and fusion of resource objects (e.g. snowflakes) typically appear as single qubit errors in the eventual TPC state. While there are instances where a two-qubit error can afflict the ultimate lattice, the majority of these involve one qubit from the prime lattice and one from the dual, i.e. the black and green qubits in Fig.~\ref{fig:topologies}b. Such correlations do not affect the fault tolerance threshold. There will be occasional instances of errors between two qubits both within the prime lattice, or both in the dual. However these are rare -- for example in the case where one uses the snowflake strategy with $p_h=0.9$, the rate for these errors is two orders of magnitude lower than the corresponding rate of single-qubit errors on the TPC (given equal rates for the various forms of error during growth)~\cite{supMat}. {\color{black} In this case, two qubit errors only weakly affect the threshold \cite{Raussendorf20062242}}, so as an approximation we can ignore such events and assume that all gate errors affect at most one qubit in each sub lattice. Thus we consider a lattice with a (low) rate of random single qubit errors, and a (relatively high) portion of missing `edges' which are known. We need to determine the threshold for such a lattice to support computation. Fortunately, our previous work has considered the closely related case of a lattice with a significant number of missing {\em nodes}. We need only map the case of missing edges to that of missing nodes in order to make contact with that analysis and thus obtain thresholds in the present case.  

Consider the standard TPC state, specifically neighbouring qubits $i$ (in the primal lattice) and $j$ (in the dual lattice). Each qubit is centred on a face of its respective sublattice, and is a member of two cubic unit cells of the sublattice.   In the ideal case where no bonds are missing, the product of cluster stabilisers associated with the faces of each cubic unit cell is simply the product of $X$ operators acting on the respective face-centred qubits, yielding two parity-check operators associated with each qubit: $P_{i}^{{1,2}}$ for qubit $i$ and $P_{j}^{1,2}$ for qubit $j$.  Since these ideal parity check operators are just products of $X$ operators on each face of the corresponding cube, they commute point wise, which enables the error syndrome to be determined by single particle $X$ measurements \cite{PhysRevLett.98.190504,Raussendorf20062242,1367-2630-9-6-199}.

In the case where the bond between qubits $i$ and $j$ is missing, the cluster stabilisers associated with the missing bond are modified.  Then, the product of cluster stabilisers centred on the cubic unit cell faces yields damaged parity check operators $\hat P_{i}^{1,2}=P_{i}^{{1,2}}Z_j$ and $\hat P_{j}^{1,2}=P_{j}^{{1,2}}Z_i$.  Whilst $\hat P_{i}^{1,2}$ and $\hat P_{j}^{1,2}$ commute, they do not commute point-wise (since $[X_i,Z_j]\neq0$).  In contrast to the ideal case, this means that determining the syndrome on the primal and dual lattices apparently requires measurement of the two-qubit operators $X_i Z_j$ and $Z_i X_j$. 

Fortunately, by simply treating the qubits $i$ and $j$ at each end of the missing bond as though they were lost, and adopting the strategy in \cite{PhysRevLett.102.200501,PhysRevA.81.022317,barrett2010fault}, we 
 form products of the damaged parity check operators, yielding \emph{super}-check operators $\tilde P_i=\hat P_{i}^{1}\hat P_{i}^{2}=P_{i}^{{1}}P_{i}^{{2}}$ and $\tilde P_j=\hat P_{j}^{1}\hat P_{j}^{2}=P_{j}^{{1}}P_{j}^{{2}}$.  These new operators are independent of the qubits $i$ and $j$,  so they are unaffected by the missing bond between them.  Furthermore each super-check operator involves only products of $X$ operators from a single sublattice, so a missing bond manifests itself as a single missing qubit on each sublattice.   This establishes a correspondence between missing bonds and correlated losses of neighbouring qubits.  Error correction is then realised by implementing the loss-tolerant, error-correcting protocol of  \cite{PhysRevLett.102.200501,PhysRevA.81.022317,barrett2010fault} to each sublattice independently \footnote{In principle, the correlated nature of the losses could be exploited using belief propagation methods \cite{PhysRevLett.104.050504}.}. 

\begin{figure}
  \centering
  \includegraphics[width=7.5cm]{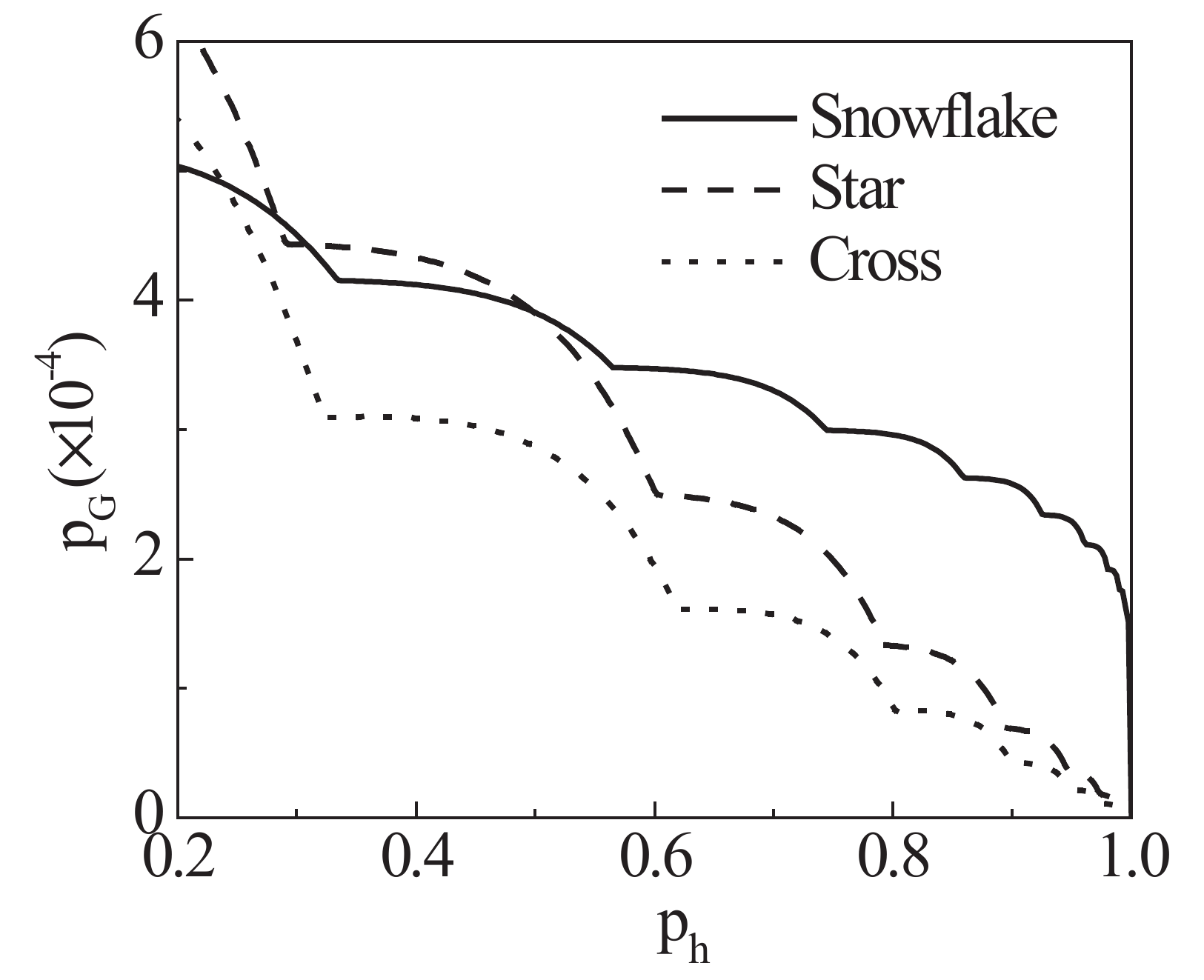}
 \caption{The principle results of our analysis. The lines define the parameter regimes where fault tolerant QIP is possible. Here for simplicity we set all gate errors, including both single-qubit and two-qubit errors, to be equally likely. This probability is denoted $p_G$ and is plotted on the vertical axis. (Memory errors due to gradual decoherence are excluded, and are shown in Fig.~\ref{fig:passiveAndSize}). Meanwhile, the probability $p_h$ of an entanglement operation failing in a heralded fashion is plotted on the horizontal axis. Note that $p_h$ can be very high, exceeding $90\%$ if the snowflake strategy is employed.  }
  \label{fig:phaseDiagram}
\end{figure}

Having made the connection to prior work on thresholds for the TPC state, we can now take any set of parameters for the low-level operations on qubits in the distributed machine, compute the effective  qubit loss rate, and determine whether quantum computation is possible. 
In Fig.~\ref{fig:phaseDiagram} we show this phase diagram under the assumption that all gate error rates are equal. We see that very high rates of heralded error can be tolerated, provided that the rate for unknown errors is below $2\times10^{-4}$. This is certainly a difficult number to achieve but might be possible in some implementations, e.g. trapped ions for which multi-qubit measurements with fidelity around this rate have already been demonstrated~\cite{PhysRevA.81.040302}.

\begin{figure}
  \centering
  \includegraphics[width=7.5cm]{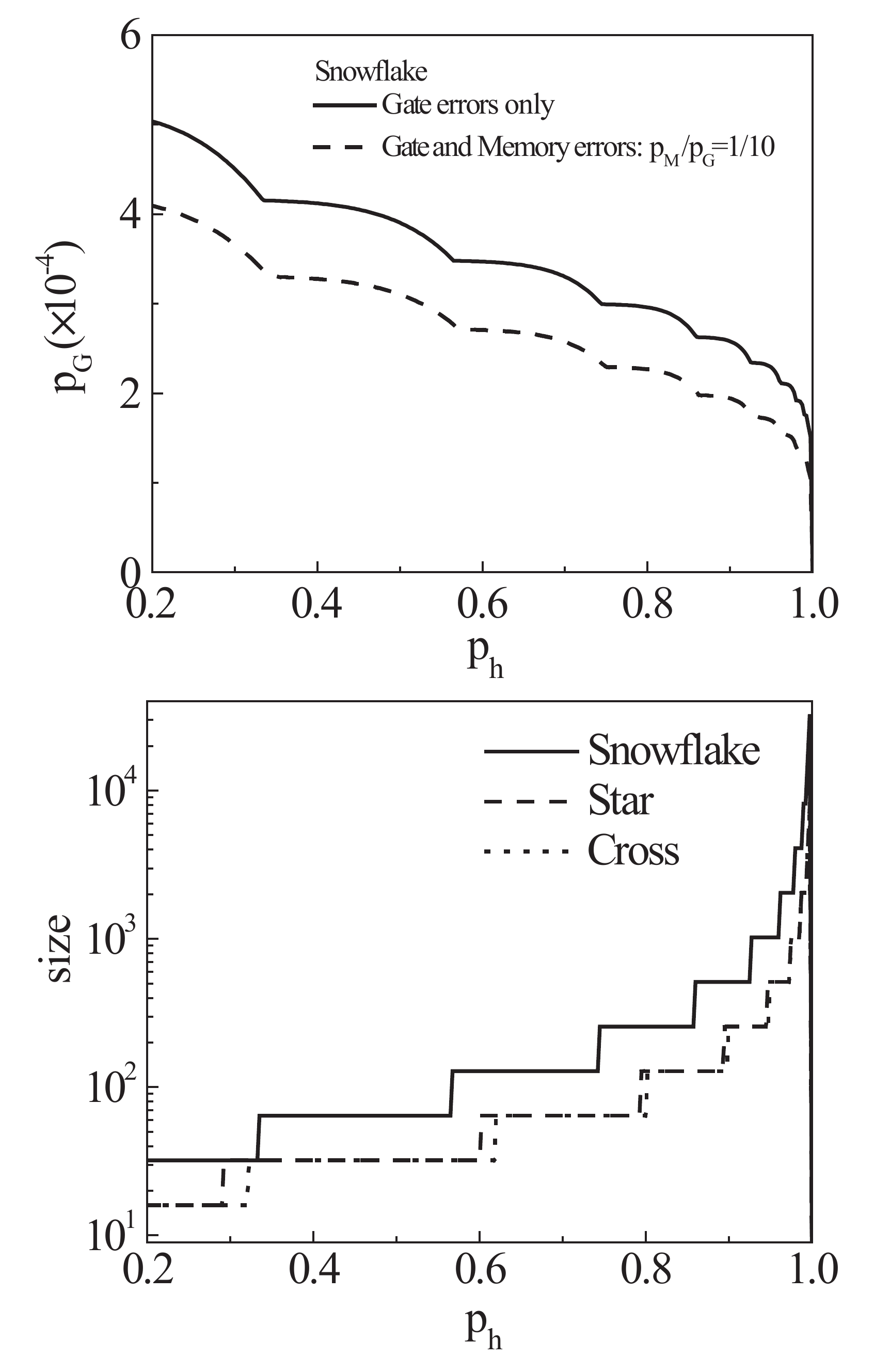}
 \caption{(a) The effect of introducing memory errors, i.e. finite gradual decoherence. (b) The size of the resource objects required to achieve the thresholds. }
  \label{fig:passiveAndSize}
\end{figure}

It remains to consider memory errors, which we assume happen at a lower rate than gate errors. In Fig.~\ref{fig:passiveAndSize}a we show the effect of `switching on' memory errors at a level equal to one tenth of the gate error rate. As one might expect, this lowers the overall threshold, but not dramatically. 

Finally we consider the question of physical resource scaling. From Fig.~\ref{fig:phaseDiagram} one might be tempted to conclude that QIP is possible with {\em extremely} high rates of heralded error, perhaps reaching $99\%$ or more. However, such a conclusion would neglect the ever increasing costs of preparing the resource objects. These objects become very large as $p_h$ approaches unity. In Fig.~\ref{fig:passiveAndSize}b we see that if $p_h$ exceeds $0.98$, the size of each snowflake must be several thousand qubits. Recall that each snowflake ultimately corresponds to a single node in the TPC state, and therefore this factor would multiply the overhead already implicit in that approach. However, values in the range of $p_h\approx0.9$ may be tenable for technologies where the individual components of the distributed computer can be mass produced.

In conclusion we have determined the threshold for quantum computation when two-qubit gates are non-deterministic. A specific case is that of a fully distributed machine, i.e. a network of components each of which contains only a single qubit. We find that it is tolerable if entanglement operations over the network fail with a rate exceeding $90\%$, provided that such failures are heralded. The tolerable rate of un-heralded errors is $2\times10^{-4}$. Our analysis should allow experimentalists to determine if single-qubit components are feasible with their particular approach, or if instead multi-qubit components must be adopted. 

\begin{acknowledgments}
\textit{Acknowledgements -} We thank Dan Browne for helpful discussions. This work was supported by the National Research Foundation and Ministry of Education, Singapore.
\end{acknowledgments}

\end{document}